\documentclass[preprint,showkeys,aps,prb]{revtex4}
\usepackage[english]{babel}

\usepackage[dvips]{graphicx}
\usepackage{amssymb}

\begin{document}


\title{
Comparison of Very Smooth Cell-Model Trajectories \\
Using Five Symplectic and Two Runge-Kutta Integrators
}

\author{Wm. G. Hoover and Carol G. Hoover \\
Ruby Valley Research Institute \\ Highway Contract 60,
Box 601, Ruby Valley 89833, NV USA \\
}
\date{\today}



\vskip 0.5cm

\begin{abstract}
Time-reversible symplectic methods, which are precisely compatible with
Liouville's phase-volume-conservation theorem, are often recommended for
computational simulations of Hamiltonian mechanics.  Lack of energy drift
is an apparent advantage of such methods. But {\it all} numerical methods
are susceptible to Lyapunov instability, which severely limits the maximum
time for which chaotic solutions can be ``accurate''. The ``advantages''
of higher-order methods are lost rapidly for typical chaotic Hamiltonians.
We illustrate these difficulties for a useful reproducible test case, the
two-dimensional one-particle cell model with specially smooth forces.  This
Hamiltonian problem is chaotic and occurs on a three-dimensional
constant-energy shell, the minimum dimension for chaos.  We benchmark the
problem with quadruple-precision trajectories using the fourth-order
Candy-Rozmus, fifth-order Runge-Kutta, and eighth-order Schlier-Seiter-Teloy
integrators.  We compare the last, most-accurate particle trajectories
to those from six double-precision algorithms, four symplectic and two
Runge-Kutta.

\end{abstract}
\maketitle

\section{Introduction}
The ongoing computational revolution in physics relies on accurate
solutions of fundamental equations, Newton's ( or Lagrange's  or
Hamilton's ) Laws of Motion, in the case of classical mechanics.
The determinism of these ordinary differential equations is illusory
in many cases, as typically the equations are ``Lyapunov
unstable''.  Such instabilities grow exponentially fast,
$\simeq e^{\lambda t}$ , where $\lambda$ is the largest Lyapunov
exponent of the solution.

Particle mechanics, our own research interest, provides many examples
ranging from one-particle chaos to biomolecule simulations using models
with many thousands of atomic degrees of freedom\cite{b1}. We consider
here the simplest  particle model for chaos, a one-body ``cell model''
with the periodic four-body cell boundaries shown in {\bf Figure 1}.
The resulting motion, approximated with the simplest possible
``leapfrog'' integrator, described below, is generally Lyapunov
unstable.\cite{b2,b3} We simplify the initial conditions by starting the
particle trajectory in the field-free cell interior.  We benchmark this
problem with three quadruple-precision integrators using timesteps chosen
to maximize accuracy. We compare the resulting benchmark trajectory to
six other trajectories from self-starting double-precision algorithms
typical of molecular dynamics simulations. Five of these algorithms are
``symplectic'', including the justifiably-popular Leapfrog
Algorithm. The two others are Runge-Kutta algorithms.

\begin{figure}
\includegraphics[width=4in]{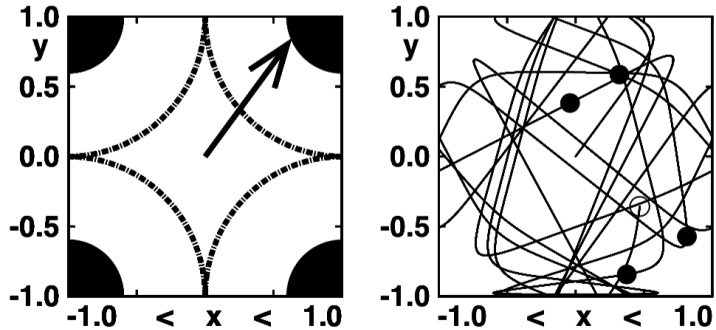}
\caption{
The periodic $2 \times 2$ unit cell is shown at the left.  The black
regions, with potential energy greater than one half, are inacessible
to the wanderer particle.  Initially the wanderer is at the origin
with velocity $(0.6,0.8)$.  Outside the central diamond-shaped region
the fixed scatterers at the cell corners exert repulsive forces on
the wanderer particle.  A visually-accurate trajectory, calculated with
a quadruple-precision fifth-order Runge-Kutta integrator, using
five million timesteps and $dt = 0.00001$, is shown at the right with
filled circles  marking the configurations at times 10, 20, 30, and 40.
The open circle corresponds to the maximum time $t=50$.
}
\end{figure}

In the following Sections we describe the specially-smooth differential
equations governing the motion of the wandering cell-model particle, and
then quantify the algorithmic accuracy with which Leapfrog and the six
more sophisticated integrators ``solve'' this same problem.  Our
conclusions make up the final Summary section.

\section{The Cell Model Trajectory in Two Space Dimensions}

Cell models played a role in models of the liquid state long before the
development of molecular dynamics.\cite{b4} The geometry treated here is
shown at the left in {\bf Figure 1}.  A mass point, the ``wanderer''
particle, moves in a periodic square cell with a motionless fixed particle
at each of the four vertices.  Using periodic boundary conditions the
equations of motion are :
$$
\dot x = (p_x/m) \ ; \ \dot y = (p_y/m) \ ; \ 
\dot p_x = F_x \ ; \ \dot p_y = F_y \ .
$$
The force on the wanderer is the gradient of the potential function $\Phi $ ,
a sum over the contributions of the four corner scatterers located at
$\{ \ r_i \ \}$ :
$$
\Phi = \sum_1^4[ \ 1 - (r-r_i)^2 \ ]^4 {\rm \ for \ } | r-r_i| < 1 \ .
$$
After advancing the coordinates one timestep $dt$ it is convenient to
localize the motion to the cell centered on the origin. Whenever the
wanderer moves ``out'', we replace it ``in'' the basic $2\times 2$
unit cell as follows  :
$$
x < -1 \rightarrow x = x + 2 \ ; \ x > +1 \rightarrow x = x - 2 \ ; \
$$
$$
y < -1 \rightarrow y = y + 2 \ ; \ y > +1 \rightarrow y = y - 2 \ .
$$
We choose initial conditions
$\{ \ x,y,p_x,p_y \ \} = \{ \ 0.0,0.0,0.6,0.8 \ \} \ $ 
and show an accurate benchmark solution of the motion equations for a
time of 50 at the righthandside of {\bf Figure 1} .  At times of 10,
20, 30, 40, and 50 the benchmark values of $(x,y)$ are :
\pagebreak

\begin{center}

10: +0.321356333887505, +0.585921713605895 \\
20: +0.81481797353866, -0.572042192203162 \\
30: -0.040449409487, +0.38290501902 \\
40: +0.3742439, -0.842854 \\
50: +0.4696, -0.3568 \\

\end{center}

{\bf Figure 1} shows a unit cell of a periodic two-dimensional lattice
in which a single particle moves in the field of scattering particles
arranged in a fixed square lattice with nearest-neighbor spacing of 2.
The potential energy maximum of unity is twice the energy of the
initial condition, shown at the center of the cell.
The benchmark solution of the motion equations $\{ \ \dot q = p \ ; \
\dot p = F(q) \ \}$ is shown at the right.  This same accurate trajectory
was obtained with both the Candy-Rozmus fourth-order and a Runge-Kutta
fifth-order integrator using 50 million and 500 million timesteps,
respectively.  The two trajectories agree throughout within visual
accuracy.  At a time of 50 $(x,y,p_x,p_y)$ are :
$$
(x,y,p_x,p_y) = (+0.46961,-0.35683,+0.11945,+0.98408 ) \ {\rm [ \ CR4 \ ]} \ ;
$$
$$      
(x,y,p_x,p_y) = (+0.46962,-0.35682,+0.11948,+0.98408 ) \ {\rm [ \ RK5 \ ]} \ .
$$ 

\section{Seven Typical Integrators and Their Trajectories}

We consider seven solution algorithms for the wanderer particle trajectory,
[1] Leapfrog (symplectic), [2] Fourth-Order Candy-Rozmus Symplectic, [3]
Monte Carlo Symplectic, [4] Sixth-Order Symplectic, [5] Fourth-Order
Runge-Kutta, [6] Fifth-Order Runge-Kutta, and [7] Eighth-Order Schlier-Seiter-Teloy
Symplectic.  For the first six of these we use a fixed timestep typical of
``accurate'' molecular dynamics simulations $dt = 0.001$ . Solutions for those
six integrators appear in {\bf Figures 2-7} . The particle mass is unity and
the energy $\Phi + K$ is one half.  For the last integrator, which has a 
trajectory visually identical to that of {\bf Figure 1}  we have chosen 
timesteps as small as 0.00000001 in order to obtain ten-digit accuracy
in the wanderer trajectory up to a time of 50 .  Let us consider the details
of all the integrators next.

\subsection{Second-Order Time-Reversible Leapfrog Algorithm}

``Symplectic'' integrators\cite{b5,b6,b7,b8,b9} automatically obey
Liouville's Theorem by advancing the solution of Hamiltonian problems
in time according to a series of phase-volume-conserving shears.
Symplectic algorithms alternate steps advancing the coordinates and
momenta in time.  The simplest example is equivalent to the
St\"ormer-Verlet ``leapfrog algorithm'' :\cite{b8,b9,b10}
$$
\{ \ {\tt q = q + (p*dt/2) \ ; \ p = p + (F/m)*dt \ ;
\ q = q +(p*dt/2)} \ \} \longleftrightarrow
$$
$$
\{ \ q_{n+1} - 2q_n + q_{n-1} \equiv  (F/m)_n \ \} \ .
$$
This algorithm is said to be ``second order''\cite{b10}, with a fixed-time
coordinate error of order $tdt^2$ for $t << 2\pi/dt^2$ when applied to the simple 
harmonic oscillator.  It is time
reversible in that changing $+dt \rightarrow -dt$ gives the same
trajectory points either forward or backward in time.

How does the simulation begin? Starting out at the origin, with the
wanderer speed equal to unity and
a fixed timestep $dt=0.001$, the first 420 steps
leave the momenta unchanged and $r^2$ becomes 1.0004. During the 421st
step the upper right scatterer
is contacted and begins to repel the wandering particle with a force :
$$
F_x = 8(x-1)(1-r^2)^3 \ ; \ F_y = 8(y-1)(1-r^2)^3 \ ; \ r = (1-x,1-y) \ ,
$$
where $x$ and $y$ are the wanderer coordinates.

After an elapsed time $t$ we reverse the sign of the time so as to
integrate {\it backward} to see how closely the wanderer returns to
its initial location. So long as $t<47$ we find that the trajectory
reverses to within a distance 0.01 of the origin.  We will see that this
retracing of steps does not guarantee a match with the accurate
trajectory shown at the right in {\bf Figure 1}. Both the trajectory
reversal and the conservation of energy are poor diagnostics for
trajectory {\it accuracy}, where accuracy means reproducing correct
values of the coordinates $x(t),y(t)$ .

\begin{figure}
\vspace{1 cm}
\includegraphics[width=2.5in,angle=-90]{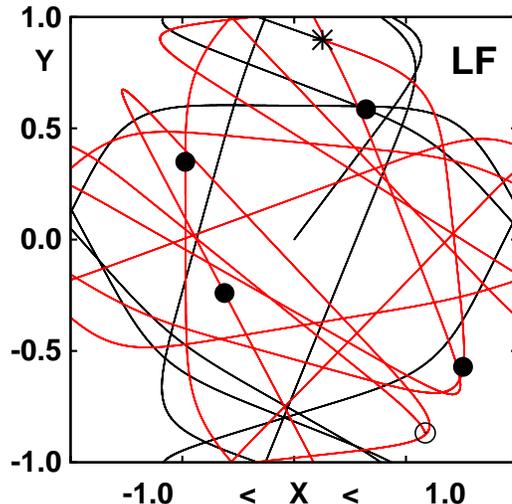}
\caption{
The Leapfrog integrator reproduces the accurate $x$ and $y$ coordinates
within 0.01 for an integration time of 18.  The energy at that point
( where the trajectory color changes, indicated by a star ) is in error
in the seventh decimal place.  Here and elsewhere the cited double-precision
times are truncated to integers because different implementations, such as
varying the order of the operations, could change these numbers.
}
\end{figure}

\subsection{Fourth-Order Time-Reversible Symplectic Integrator}

Higher-order algorithms, with fixed-time integration errors of order
$dt^3$, $dt^4$, $dt^5$ \dots  can be developed from Taylor's series
about $t$ giving small increments in the coordinates and momenta as
three-, four-, five- \dots term series in $dt$ .  Candy and
Rozmus' fourth-order integrator (with an error of order $dt^4$ at
a fixed not-too-large, time) is a simple example, cited in the
very useful summary paper by Gray, Noid, and Sumpter\cite{b7} :
$$
{\tt q = q + 0.6756036p*dt \ ; \ p = p + 1.3512072(F/m)*dt \ ; }
$$
$$
{\tt q = q - 0.1756036p*dt \ ; \ p = p - 1.7024144(F/m)*dt \ ;
\ q = q - 0.1756036p*dt \ ;} 
$$
$$
{\tt p = p + 1.3512072(F/m)*dt \ ; \ q = q + 0.6756036p*dt \ .}
$$
Reference 7 gives the analytic forms of all of the coefficients.
Notice that the coefficients incrementing the coordinates sum to 
unity as do also those incrementing the momenta.
Each timestep requires three separate evaluations of the forces.

\begin{figure}
\vspace{1 cm}
\includegraphics[width=2.5in,angle=-90]{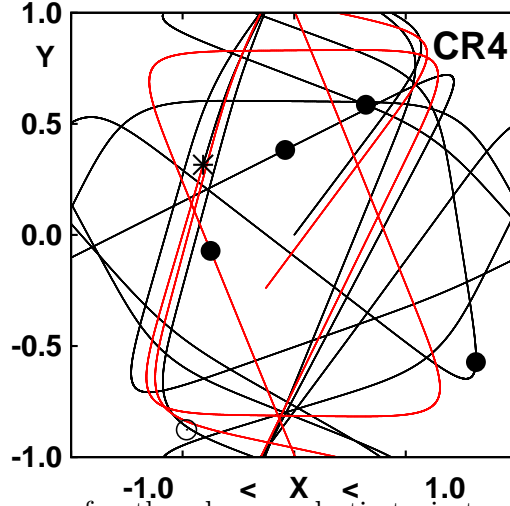}
\caption{
The Candy-Rozmus fourth-order symplectic trajectory exhibits a
color change at a time of 34, the maximum for which the coordinate
errors are less than 0.01 .  The energy error at that time is in
the twelfth decimal place.  The maximum time at which a reversed
trajectory returns to the origin within 0.01 is $t = 42$ .
}
\end{figure}

\subsection{Monte-Carlo Time-Reversible Symplectic Integrator}

Although it is usual to provide coefficients in integration
algorithms to many significant figures, in most cases an approximate
rendition is sufficient.  It is quite possible to develop algorithms
with a Monte Carlo method, adjusting the coefficients to minimize
the trajectory error for the simple harmonic oscillator problem. An
integrator requiring five force evaluations per timestep was developed
by Monte Carlo sampling\cite{b6} adjusting the coefficients subject to
the constraints of time reversibility and normalization so that the
Monte Carlo trajectory optimization occurs in a four-dimensional space.
The resulting integrator was successful in modelling many-body dynamics
but is here applied to the cell-model problem of {\bf Figure 1} :
$$
{\tt q = q + 0.005904p*dt \ ; \
p = p + 0.171669(F/m)*dt \ ;}
$$
$$
{\tt q = q + 0.5l5669p*dt \ ; \
p = p - 0.516595(F/m)*dt \ ;}
$$
$$
{\tt q = q - 0.021573p*dt \ ; \
p = p + 1.689852(F/m)*dt \ ; \
q = q - 0.02l573p*dt \ ;}
$$                                                                            $$
{\tt p = p - 0.516595(F/m)*dt \ ; \
q = q + 0.5l5669p*dt \ ;}
$$                                                                            $$
{\tt p = p + 0.171669(F/m)*dt \ ; \
q = q + 0.005904p*dt \ .}
$$
The cell model trajectory using this Monte Carlo integrator is illustrated in
{\bf Figure 4}.  

\begin{figure}
\vspace{1 cm}
\includegraphics[width=2.5in,angle=-90]{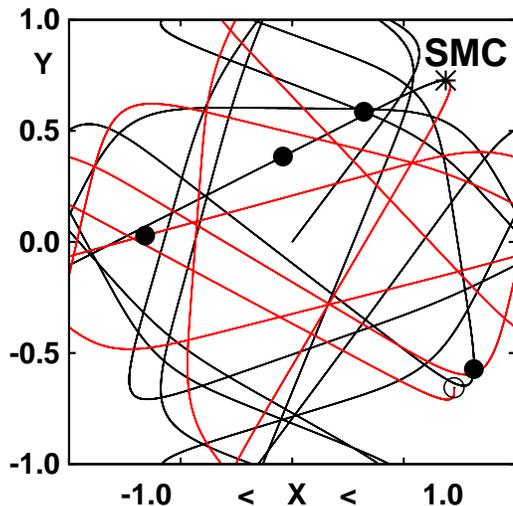}
\caption{
The color change in the trajectory from the Monte Carlo symplectic integrator
occurs at a time of 31, after which the coordinate errors exceed 0.01.  The
energy error there is in the thirteenth digit.  For this integrator a trajectory
reversed at a time of 43 will return to the origin with coordinates recurring
within 0.01.
}
\end{figure}

\subsection{Yoshida's Sixth-Order Time-Reversible Integrator}

Yoshida developed and applied a general technique for finding a variety of
higher-order symplectic integrators.\cite{b11}  His sixth-order time-reversible
integrator advances the coordinates $(\Delta q \propto pdt)$ eight times per
timestep, using the symmetric (so as to guarantee time-reversibility) set of eight
coefficients which sum to unity :
$$
+0.39225680523878, +0.51004341191846,-0.47105338540976,+0.06875316825252,
$$
$$
+0.06875316825252,-0.47105338540976,+0.51004341191846,+0.39225680523878.
$$
Between the successive coordinate updates there is a force calculation
and an update of the momenta $(\Delta p \propto Fdt)$, using seven
coefficients, which likewise sum to unity :
$$
+0.78451361047756,+0.23557321335936,-1.1776799841789,+1.3151863206839,
$$
$$
-1.1776799841789,+0.23557321335936,+0.78451361047756.
$$

\begin{figure}
\vspace{1 cm}
\includegraphics[width=2.5in,angle=-90]{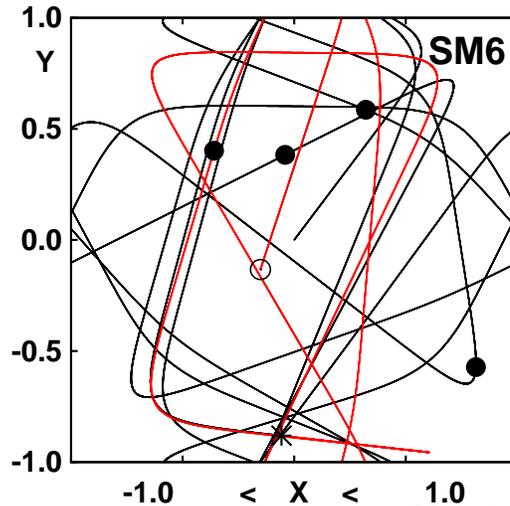}
\caption{
This double-precision trajectory is based on Yoshida's time-reversible
sixth-order integrator with a timestep $dt = 0.001$ .  There is a color
change at $t=36$ , indicating the degradation of trajectory accuracy
to $\pm0.01$ despite the negligible energy error in the fourteenth
decimal place.  Trajectory reversal at a time of 42 returns to the
origin within coordinate errors of 0.01.
}
\end{figure}

\subsection{Fourth-Order and Fifth-Order Runge-Kutta Integrators}

Runge-Kutta integrators ( {\it circa} 1900, as described in Wikipedia )
advance both coordinates and momenta {\it simultaneously} in a series of
stages within each timestep $dt$.  As in the symplectic case the variables
at time $t+dt$ are expressed as series in $dt$, putting conditions on the
summed-up coefficients for each power of $dt$ to be treated correctly by
the algorithm.

The main advantage of Runge-Kutta methods is that they can be applied to
arbitrary sets of ordinary differential equations, not just those from
Hamiltonian mechanics.  The fourth-order ``classic'' Runge-Kutta method
has been a standard workhorse model for solving sets of coupled ordinary
differential equations for 100 years.  Applied to the harmonic oscillator
the fourth-order algorithm suffers a loss in energy proportional to the
fifth power of the timestep.  The  fifth-order Runge-Kutta integrator
behaves in the opposite manner with the energy increasing rather than
decreasing.

\begin{figure}
\vspace{1 cm}
\includegraphics[width=2.5in,angle=-90]{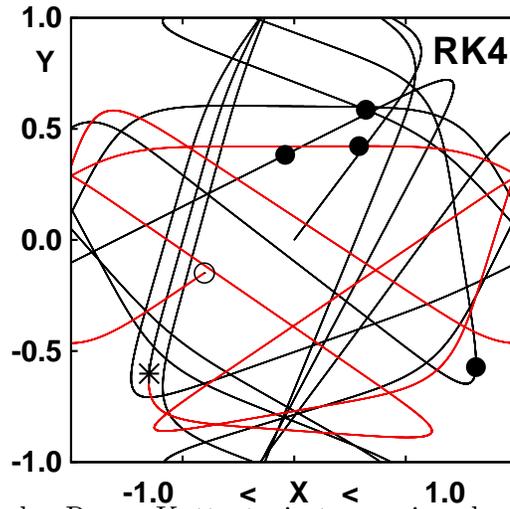}
\caption{
The fourth-order Runge-Kutta trajectory using double precision and a
timestep $dt  = 0.001$ provides coordinates accurate within 0.01 through
a time of 35, indicated by the color change at the star.  The energy
error at that point, $10^{-13}$, is negligible.  Changing the sign of
the timestep at $t=42$ , $+dt \rightarrow -dt$ , returns the trajectory
to the origin within a precision of 0.01 .
}
\end{figure}

\begin{figure}
\vspace{1 cm}
\includegraphics[width=2.5in,angle=-90]{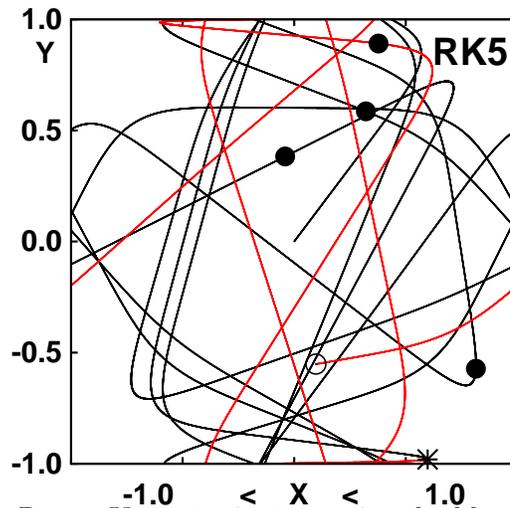}
\caption{
The fifth-order Runge-Kutta trajectory using double precision and a
timestep $dt  = 0.001$ provides coordinates accurate within 0.01 through
a time of 37, indicated by the color change at the star.  The energy
error at that point is $10^{-14}$ .  For times less than 42 reversing
the trajectory, by
setting $+dt \rightarrow -dt$ , returns the trajectory to the origin with
precision 0.01.  This integrator is the best of the double-precision
integrators tested here.  Any one of the five higher-order integrators
is accurate for about twice the time of the second-order Leapfrog
integrator.
}
\end{figure}
Hybrid ``adaptive'' models, incorporating both fourth- and fifth-order
algorithms, provide a simple means for the automatic control of
integration errors.  The harmonic oscillator is an excellent test case
of integrator accuracy where Lyapunov instability is absent.\cite{b10}
{\bf Figure 6} illustrates the same cell-model orbit for the classic
fourth-order Runge-Kutta integrator.  {\bf Figure 7} shows a fifth-order
Runge-Kutta integrator :
\pagebreak
\noindent
$$
\noindent
{\tt yp1 = yp[ \ y \ ]}
$$
\noindent
$$
{\tt yp2 = yp[ \ y + (dt/2)*yp1 \ ]}
$$
\noindent
$$
{\tt yp3 = yp[ \ y + (dt/16)*(3yp1 + yp2) \ ]}
$$
\noindent
$$
{\tt yp4 = yp[ \ y + (dt/2)*yp3 \ ]}
$$
\noindent
$$
{\tt yp5 = yp[ \ y + (dt/16)*(-3yp2+6yp3+9yp4) \ ]}
$$
\noindent
$$
{\tt yp6 = yp[ \ y + (dt/7)*(yp1+4yp2+6yp3-12yp4+8yp5) \ ]}
$$
\noindent
$$
{\tt y = y  + (dt/90)*(7yp1+32yp3+12yp4+32yp5+7yp6)}
$$
Here {\tt yp[ \ ... \ ]} represents the righthandside of the vector
differential equation $\dot y = y'$ where the six force evaluations in
each timestep are indicated by $\{ \ yp1, \ yp2, \ \dots \ yp6 \ \}$ .

Both Runge-Kutta integrators return to the origin with errors no more
than 0.01 with reversal at time 42.  Forward in time their trajectories
are accurate through times of 35 and 37, the last being the best of the
double-precision integrators.  The energy errors for the two Runge-Kutta
integerators are in the thirteenth and fourteenth decimal places.

\subsection{An Eighth-Order Time-Reversible Symplectic Integrator}

Ernst Teloy, Christoph Schlier, and Ansgar Seiter developed and implemented a
useful eighth-order time-reversible symplectic integrator with 17 force
evaluations per step.  Applied to the harmonic oscillator the rms coordinate error
increases by about eight orders of magnitude when the timestep is increased by
a factor of ten, consistent with an eighth-order method.

For the reader's convenience we reproduce here the 18 coefficients required to
implement the method.  They can be found quoted to 35 decimal places at
Christoph Schlier's Freiburg website or in Reference 12 . This precision is
steadily reduced, digit by digit, through Lyapunov instability, described in
more detail in Section IV.  In the cell-model case the rate of precision loss
is 0.7 , one binary bit per unit time.  Accordingly, for the eighth-order
integrator in quadruple precision at time 50 we would expect an exponentially
amplified error of order $10^{-32} \times 2^{50} \simeq 10^{-17}$ . In fact, we
find a trajectory error of order $10^{-10}$ using a timestep of $10^{-8}$ , as
is shown below.

Even so the eighth-order integrator with $dt = 0.001$ loses only seven of the
original 35 digits in energy along with twenty digits in position when run
forward and backward for 50,000 steps to match the time illustrated in all the
Figures. As was illustrated and emphasized in References 12 and 13 energy
conservation and trajectory reversibility are both of them misleading diagnostics
of trajectory accuracy. It is only through a study of {\it convergence} that
trajectories can be validated.  For the eighth-order symplectic integrator the
timestep dependence of the $(x,y)$ coordinates at time 50 is as follows :
$$                                                                                           
dt = 0.00100000 \rightarrow (0.48704 \ 51729, \  +0.13435 \ 10401 )                          
$$
$$                                                                                           
dt = 0.00010000 \rightarrow ( 0.48185 \ 80396, \ -0.32559 \ 07485 )                          
$$
$$                                                                                           
dt = 0.00001000 \rightarrow (0.46961 \ 32018, \ -0.35683 \ 11339 )                           
$$
$$                                                                                           
dt = 0.00000100 \rightarrow (0.46961 \ 40145, \ -0.35682 \ 95856 )                           
$$
$$                                                                                           
dt = 0.00000010 \rightarrow (0.46961 \ 40143, \ -0.35682 \ 95861 )                           
$$
$$                                                                                           
dt = 0.00000001 \rightarrow (0.46961 \ 40142, \ -0.35682 \ 95862 )                           
$$
For the convenience of the reader we reproduce the integrator
coefficients here from Reference 12, together with a short harmonic-oscillator
program to demonstrate their use.
\pagebreak
\subsection{Schlier-Seiter-Teloy Integrator Coefficients}
\noindent
{\tt
       c( 1) = +0.04463 79505 23590 22755 91399 96257 33590 d00 \\
       c( 2) = +0.13593 25807 16909 59145 54326 42134 95574 d00 \\
       c( 3) = +0.21988 44042 71470 72254 44553 50696 06167 d00 \\
       c( 4) = +0.13024 94678 05238 28601 62119 37781 96846 d00 \\
       c( 5) = +0.10250 36569 39750 69608 26124 10077 79814 d00 \\
       c( 6) = +0.43234 52186 93585 47487 98325 78848 77035 d00 \\
       c( 7) = -0.00477 48291 69168 81658 02248 90639 62934 d00 \\
       c( 8) = -0.58253 47690 40408 45493 11283 79308 61212 d00 \\
       c( 9) = -0.03886 26428 21118 17697 73742 08751 89743 d00 \\
       c(10) = +0.31548 72853 79404 79698 27360 37972 74199 d00 \\
       c(11) = +0.18681 58374 32971 55471 52615 35039 72746 d00 \\
       c(12) = +0.26500 27549 90620 83398 34600 29630 79872 d00 \\
       c(13) = -0.02405 08473 57473 61993 57358 79824 07554 d00 \\
       c(14) = -0.45040 49249 97722 51180 92289 67121 51891 d00 \\
       c(15) = -0.05897 43301 55923 86914 57532 39267 66330 d00 \\
       c(16) = -0.02168 47617 18613 35324 93438 86847 07580 d00 \\
       c(17) = +0.07282 08003 35901 28173 76189 26412 34244 d00 \\
       c(18) = +0.55121 42963 41970 67334 40560 13815 94315 d00 \\
}
Oscillator program with {\tt q,p,dt} and 18 {\tt c(i)} :    \\
{\tt
       do i = 1,17,2                       \\
       q = q + c(i)*p*dt                   \\
       p = p - c(i+1)*q*dt                 \\
       enddo                               \\
       do i = 17,1,-2                      \\
       q = q + c(i)*p*dt                   \\
       if(i.gt.1) p = p - c(i-1)*q*dt      \\
       enddo
}

\section{Lyapunov Instability in the Cell Model}

For chaotic systems the algorithmic accuracy of numerical integrators
deteriorates exponentially rather than linearly in the time\cite{b3}.
The underlying exponential Lyapunov instability of dynamical systems is easily
measured by following the motion of a ``reference'' trajectory in the usual
way, for instance with any one of the seven algorithms discussed here.  An
additional ``satellite'' trajectory, separated from the reference by a
small length $\delta_0$ , is also followed using the same algorithm. At the
end of each timestep the separation is rescaled, maintaining the length of
the offset between the trajectories constant, but allowing the direction to
vary :
$$
\delta(t+dt) \equiv [ \ r_s(t+dt) - r_r(t+dt) \ ] \ ; \
$$
$$
r_s \longrightarrow r_r(t+dt) + \delta(t+dt)[ \ \delta_0/| \ \delta(t+dt) \ | \ ] \ .
$$
The largest Lyapunov exponent is simply the average value of the growth
rates measured at the ends of every timestep prior to rescaling :
$$
\lambda_1 = \langle \ (1/dt)\ln[ \ | \ \delta(t+dt) \ |/\delta_0 \ ] \ \rangle \ .
$$
Previous studies of this cell model,\cite{b3} with the same initial condition,
have shown that the largest Lyapunov exponent is about 0.7.  This means that an
error of the order $10^{-16}$ at the initiation of a run of length 50 will
increase by a factor of
$e^{\lambda t} = e^{0.7 \times 50} = e^{35} \simeq 10^{15} \ $.

This exponential growth rate explains why it is that {\it all} of the
double-precision integrators fail, from the standpoint of reproducing a
reversible trajectory, at about the same time, at  about {\it half} the
time where quadruple-precision trajectories fail.  It is because these
trajectories are just approximations that the most sophisticated
biomolecule simulations are based on the rudimentary leapfrog algorithm
rather than more sophisticated algorithms.

Of
course, even the slightest difference in the error prior to amplification
will yield a different history.  Just summing the
particle interactions in a different order leads to qualitatively
different histories once the Lyapunov instability rises to the level of
visibility, an increase of 16 digits for routine double-precision
simulations.  The phase-shift errors in all of the algorithms discussed
here can be measured by choosing the initial velocity
$(\sqrt{1/2},\sqrt{1/2})$ for which the roundoff errors in the $x$ and $y$
directions are identical.

If high accuracy is required, as in astronomical simulations, multiple
precision can be employed, as demonstrated by Lorenz Attractor simulations
using a precision of thousands of decimal digits.  But, as Joseph Ford was
fond of pointing out, Lyapunov instability is incompatible with high
accuracy.  Doubling the number of significant figures in the integration
algorithm only doubles the time for which the simulation is accurate.

Recently Hanno Rein and David Siegel\cite{b14} developed and implemented
a relatively complicated fifteenth-order  integrator for gravitational
problems with the provocative title ``A Fast, Adaptive, High-Order
Integrator for Gravitational Dynamics, Accurate to Machine Precision
Over a Billion Orbits''.  Evidently this integrator is not at all intended
for long-time applications to chaotic problems, where errors grow
exponentially with time.  Conversations with Ben Leimkuhler and Mark
Tuckerman, both of whom summarily dismiss the use of Runge-Kutta
techniques, due to their monotonic energy drift, plus the appearance of
Rein and Siegel's high-order long-time work led to the present article.

\section{Recent Developments and Summary}

To summarize, for simple chaotic simulations ( such as classical fluids )
symplectic integrators attain accuracies similar to those obtained with
Runge-Kutta integration and are primarily limited by Lyapunov instability.
Although energy conservation and trajectory reversibility characterize
symplectic integrators, those properties do not ensure trajectory accuracy.
The reversibility of the double-precision leapfrog integrator, to a time
of 47 and back, exceeds that of all the more accurate double-precision
integrators.

For us it was illuminating to find that the humble Leapfrog integrator,
presumably nearing its 330th anniversary\cite{b8}, is nearly as useful
as are its more complex relatives, and is certainly far more economical.
For higher accuracy there is little distinction between the symplectic
and the Runge-Kutta integrators for chaotic problems, because both types
lose accuracy at the very same rate, determined by the maximum Lyapunov
exponent. 

It is significant that all of the integrators used here conserve energy
almost perfectly for the benchmark problem.  They also reverse back to
the initial conditions {\it even when their trajectories are inaccurate}.
One takeaway message from these simulations is the one to which Joseph Ford
devoted much thought and many thought-provoking words, among them these 
taken from Reference 16 :

\begin{quote}
``Newtonian determinism assures us that chaotic orbits exist and are
unique, but they are nevertheless so complex that they are humanly
indistinguishable from realisations of truly random processes.''
\end{quote}

Liao has confronted the Lyapunov instability problem headon for the Lorenz
Attractor.\cite{b17} By using 3500-term series expansions coupled with
4180-digit arithmetic he followed the evolution of the Lorenz Model to
a time of 10,000.  Like the continuing discovery of the digits of $\pi$
this activity will last as long as mankind.

Lyapunov instability often shows up in peculiar places.  Simply changing
the order of operations in adding up forces or in computing the weights
of contributions to differential equations' righthandsides can provide the
seeds from which macroscopic change develops.  We learned this lesson in
simulating the collisions of mirror-image manybody drops and crystals.  To
retain accurate mirror symmetry it was necessary to symmetrize the force
calculations at every timestep.\cite{b18}  

\section{Acknowledgments}

We thank Ben Leimkuhler, Hanno Rein, and Mark Tuckerman for their useful
and stimulating comments, Clint Sprott for his constant encouragement,
and Christoph Schlier for several helpful emails.
We are grateful to the GNU Fortran project for furnishing their no-cost
quadruple-precision compiler.  To find it GOOGLE ``GNU Fortran project''.


\begin{thebibliography}{99}


\bibitem{b1}   M. Karplus, `` \ `Spinach on the Ceiling' : a Theoretical 
               Chemist's Return to Biology'', Annual Review of
               Biophysics and Biomolecular Structure {\bf 35}, 1-47
               (2006).

\bibitem{b2}   H. A. Posch, W. G. Hoover, and F. J. Vesely, ``Canonical
               Dynamics of the Nos\'e Oscillator: Stability, Order, and
               Chaos'', Physical Review A {\bf 33}, 4253-4265 (1986).

\bibitem{b3}   W. G. Hoover and C. G. Hoover, {\it Simulation and Control
               of Chaotic Nonequilbrium Systems} (World Scientific
               Publishers, Singapore, 2015). 

\bibitem{b4}   J. A. Barker and D. Henderson, ``What is `Liquid' ? 
               Understanding the States of Matter'', Reviews of Modern
               Physics {\bf 48}, 587-671 (1976).

\bibitem{b5}   B. Leimkuhler and S. Reich, {\it Simulating Hamiltonian
               Dynamics} (Cambridge University Press, United Kingdom, 2004).

\bibitem{b6}   W. G. Hoover, O. Kum, and N. E. Owens [ now Nancy Fulda ],
               ``Accurate Symplectic Integrators {\it via} Random
               Sampling'', the Journal of Chemical Physics {\bf 103},
               1530-1532 (1995).

\bibitem{b7}   S. K. Gray, D. W. Noid, and B. G. Sumpter, ``Symplectic
               Integrators for Large Scale Molecular Dynamics Simulations:
               A Comparison of Several Explicit Methods'', the Journal of
               Chemical Physics {\bf 101}, 4062-4072 (1994).

\bibitem{b8}   E. Hairer, C. Lubich, and G. Wanner, ``Geometric Numerical
               Integration Illustrated by the St\"ormer-Verlet Method'',
               Acta Numerica {\bf 12}, 399-450 (2003).

\bibitem{b9}   D. Levesque and L. Verlet, ``Molecular Dynamics and Time
               Reversibility'', Journal of Statistical Physics {\bf 72},
               519-537 (1993).

\bibitem{b10}  G. D. Venneri and W. G. Hoover, ``Simple Exact Test for
               Well-Known Molecular Dynamics Algorithms'', Journal of
               Computational Physics {\bf 73}, 468-475 (1987).

\bibitem{b11}  H. Yoshida, ``Construction of Higher Order Symplectic 
               Integrators'', Physics Letters A {\bf 150}, 262-268 (1990).

\bibitem{b12}  Ch. Schlier and A. Seiter, ``High-Order Symplectic Integration: An
              Assessment'', Computer Physics Communications {\bf 130}, 176-189
              (2000).

\bibitem{b13}  Ch. Schlier and A. Seiter, ``Symplectic Integration of Classical               
              Trajectories: A Case Study'', Journal of Physical Chemistry
              {\bf  102}, 9399-9404 (1998).

\bibitem{b14}  H. Rein and D. S. Spiegel, ``IAS15: A Fast, Adaptive,
               High-Order Integrator for Gravitational Dynamics, Accurate
               to Machine Precision Over a Billion Orbits'', ar$\chi$iv
               1409.4779, also available in the Monthly Notices of the
               Royal Astonomical Society. 

\bibitem{b15}  W. G. Hoover, J. C. Sprott, and P. K. Patra, ``Ergodic
               Time-Reversible Chaos for Gibbs' Canonical Oscillator'',
               (submitted to Physical Review E, 2015) = ar$\chi$iv
               1503.06729.

\bibitem{b16}  J. Ford, ``What is Chaos, that We Should be Mindful of
               I≈ß≈çt ?'', in {\it The New Physics}, edited by Paul Davies
               (Cambridge University Press, 1989).

\bibitem{b17}  S. Liao, ``A Comment on the Arguments about the
               Reliability and Convergence of Chaotic Simulations'',
               International Journal of Bifurcation and Chaos, {\bf 24},
               91450119 (2014) = ar$\chi$iv 1401.0256.

\bibitem{b18}  W. G. Hoover and C. G. Hoover, ``What is Liquid? Lyapunov
               Instability Reveals Symmetry-Breaking Irreversibility
               Hidden within Hamilton's Many-Body Equations of Motion'',
               Condensed Matter Physics {\bf 18}, 13003-1-13 (2015) =
               ar$\chi$iv 1405.2485.
\end{thebibliography}
\end{document}